\documentclass[11pt,a4paper]{article}
\usepackage{jheppub_kim}

\usepackage{pdflscape}
\usepackage{amsmath}
\usepackage{amssymb}
\usepackage{dcolumn}
\usepackage{bm}
\usepackage{color}
\usepackage{epsfig}
\usepackage{amsfonts}
\usepackage{graphicx}
\usepackage{subfigure}
\usepackage{dcolumn}

\newcommand{\be}{\begin{equation}}
\newcommand{\ee}{\end{equation}}
\newcommand{\bea}{\begin{eqnarray}}
\newcommand{\eea}{\end{eqnarray}}

\def\be{\begin{equation}}
\def\ee{\end{equation}}
\def\bea{\begin{eqnarray}}
\def\eea{\end{eqnarray}}

\begin{document}
\title{Quantum gravity effects on Ho\v rava-Lifshitz black hole}
\author[a]{B. Pourhassan,}
\author[b]{S. Upadhyay,}
\author[c]{H. Saadat,}
\author[a]{and H. Farahani,}

\affiliation[a]{School of Physics, Damghan University, Damghan, 3671641167, Iran}
\affiliation[b]{Centre for Theoretical Studies, Indian Institute of Technology Kharagpur,  Kharagpur-721302, WB, India}
\affiliation[c]{Department of  physics, Shiraz Branch, Islamic Azad University, Shiraz,
P. O. Box 71555-477, Iran}

\emailAdd{b.pourhassan@du.ac.ir}
\emailAdd{sudhakerupadhyay@gmail.com}
\emailAdd{hsaadat2002@yahoo.com}
\emailAdd{h.farahani@umz.ac.ir}

\abstract{In this paper, we would like to obtain quantum gravity effects by using Ho\v rava-Lifshitz black hole. We consider logarithmic corrected thermodynamics quantities and investigate the effects of logarithmic correction term. Logarithmic correction comes from thermal fluctuation and may be interpreted as quantum loop corrections. As black hole is a gravitational system, hence we can investigate quantum gravity effect. We find such effects on the black hole stability and obtain domain of correction coefficient.}


\maketitle

\section{Overview and motivation}
As we know, black holes are related to the usual thermodynamics laws and   are objects with maximum entropy \cite{1,2,3}. In fact, the study of black hole thermodynamics is a  subject of considerable
physical importance in
the development of quantum field theory in curved space-time.
It is found that the black hole entropy $(S_{0})$ depends on the black hole event horizon area ($A$) rather than the black hole volume ($V$) with initial formula,
\begin{equation}\label{00}
S_{0}=\frac{A}{4}.
\end{equation}
The  black hole thermodynamics plays an important role in, for example,
the very early universe.
Black hole thermodynamics provides a real connection between gravity and quantum mechanics
 \cite{page}. In this connection, there are  several kinds of black holes have been studied
 and black holes with lower dimensions are   really been interesting. For instance, $2D$ black holes with vanishing horizon area are discussed in Refs. \cite{2d1,2d2,2d3,2d4,2d5}. Thermodynamics of different black holes with hyperscaling violation are  studied in the Refs. \cite{hyper1,hyper2}. Recently, the thermodynamics, stability and   Hawking-Page phase transition of the massive Banados, Teitelboim, and Zanelli (BTZ) black holes  with energy dependent space-time are studied \cite{sud}.
A full discussion on getting   the thermodynamic and statistical mechanical properties of black holes is given in  \cite{car}.
 A brief review on black hole thermodynamics can be found in Ref. \cite{Ross}, where the Unruh effect and
Hawking radiation, showing how quantum fields on black hole backgrounds behave
thermally are discussed. Here, a discussion
of entropy from the Euclidean path-integral point of view and the understanding of
 black hole thermodynamics in AdS/CFT can also be found.

Despite knowing the fact that the black holes much larger than the Planck scale have entropy proportional to its
horizon area, it is important to investigate  the leading order corrections to entropy, as one reduces the size of the black hole.  These corrections interpreted as quantum effect,  due to quantum fluctuations,  in turn  modify the holographic principle \cite{6b,6a}.
  Kaul and Majumdar \cite{kaul} derived the lowest order
corrections to the Bekenstein-Hawking entropy in a particular formulation  of the
 quantum geometry  program of Ashtekar et al and found that the leading correction is
logarithmic, with
\begin{equation}\label{001}
S=  S_{0} + \alpha \log A+\cdots
\end{equation}
where coefficient $\alpha$ depends on the details of the model and dots denotes higher order corrections. The leading order correction  to the geometry of large AdS black holes in a   four-dimensional Einstein gravity with
negative cosmological constant is discussed \cite{th} and found that the Hawking temperature grows without bound with increasing black hole mass.  It may be noted that corrections to the thermodynamics of black holes can be studied using the non-perturbative quantum  general relativity. It is also possible to study the corrected  thermodynamics of
a black hole with the help of matter fields in neighborhood of a black hole \cite{other, other0, other1}.

The study of  logarithmic corrections in various contexts are subject of current interests
\cite{sud1,godel,1503.07418,1505.02373}.
For example, it has been   studied  in the contexts of  G\"{o}del black hole \cite{godel},
 Schwarzschild-Beltrami-de Sitter black hole \cite{sud2} and   massive black hole in  AdS space
 \cite{sud3}.  The  corrected thermodynamics of a  dilatonic black hole has also been studied   \cite{jy} to find the same universal form of correction term. The partition
function of a black hole  is used to study the corrected thermodynamics
of a black hole \cite{bss}. The universality of the correction can be understood in the Jacobson formalism, where the  Einstein
equations are famous thermodynamics identities \cite{z12j, jz12}. As a result,
the quantum correction
to the structure of space-time would produce thermal fluctuations in the black holes thermodynamics.
Such corrected thermodynamics has  same universal shape as expected from the quantum gravitational effects \cite{l1, SPR, more}. Therefore, the equation (\ref{001}) can be extended to the following forms,
\begin{equation}\label{002}
S=  S_{0} +\alpha \log [S_{0}T_H^{2}]+\cdots  =  S_{0}+\alpha  \log [C_{0}T_H^{2}]+\cdots,
\end{equation}
where $C_{0}$ is uncorrected specific heat. In that case quantum corrected charged dilaton 2d black holes and BTZ black holes has been studied already \cite{Odi1, Odi2, Odi3}.
 Thermodynamics of a small singly spinning Kerr-AdS black hole under the effects of thermal fluctuations has been studied by the Ref. \cite{NPB} and concluded that this form of corrections (logarithmic correction) becomes important when the size of the black hole is sufficient small.   Such corrections may affect the critical behaviors of black object, for example in the Ref. \cite{PRD} a dyonic charged AdS black hole (holographic dual of a
van der Waals fluid) considered and logarithm-corrected thermodynamics investigated to show that holographic picture is still valid.\\
On the other hand, Ho\v rava-Lifshitz (HL) gravity   theory,  a new approach to deal the quantum gravity,    is based on the idea of the breaking
of the Lorentz invariance by equipping the space-time with additional geometric structure  \cite{4}. The HL gravity has been studied widely  \cite{5,6,7,8}, and  has been considered in several works of particle physics and cosmology \cite{dark, CJP}.  The static
spherically symmetric black hole solutions in  HL theory is investigated in Ref. \cite{12}. The generalized second law of thermodynamics in Ho\v rava-Lifshitz cosmology investigated by \cite{kise}.
The thermodynamics of HL black holes has been studied  in \cite{main, 9, 10, 11},  which has certain instabilities. The thermal fluctuation (quantum gravity effects) for HL black hole
is still to explore.  We would like this opportunity to  explore the quantum gravity effects on the thermodynamics of HL gravity due to the first order correction.

In this paper, we consider HL black hole with Lu-Mei-Pop (LMP) solution
and discuss  the effects of thermal fluctuations to the entropy of a black hole which gets first-order (logarithmic)
correction, and it is interpreted as quantum effects because these are important as the size
of black hole reduces due to the Hawking radiation. We discuss these thermal fluctuations
for the curvatures corresponding to spherical, flat and hyperbolic horizon.
We find that the first order-corrected equations of states also satisfy the
first-law of thermodynamics for all three curvatures. In order to
see the effects of thermal fluctuations, we do comparative analysis  of
corrected thermal quantities with uncorrected ones. Here, in case of spherical space, we find that the pressure with negative correction coefficient shows opposite behavior
when horizon radius tending to zero and pressure takes positive value for large
value of radius horizon. However, in case of flat space, the pressure is
 negative for small black hole only when the higher positive values of correction parameter.
 In case of hyperbolic space,   for both the cases of corrected and uncorrected one, the pressure is increasing
function and takes positive value for finite size of black hole. However, the pressure is
decreasing function for very small horizon radius in both the uncorrected and corrected with positive  coefficients and corrected pressure becomes asymptotically negative when horizon radius tends to zero.
 The Helmholtz free energy for spherical space takes negative asymptotic value for
 negative correction parameter only when horizon radius tends to zero.   There exists a critical radius of horizon for such black hole. With higher positive value of
 correction parameter the Helmholtz free energy falls faster till  critical horizon radius
and then falls  bit slower. In flat space, the Helmholtz free energy for black hole   is an increasing function always. In hyperbolic space, the behavior of the Helmholtz free energy
is negative to that of pressure. In spherical space,  there exists a critical radius for HL black hole and
the large black hole   without thermal fluctuations is in completely
stable phase, however the  small black hole   is in completely unstable phase.
For hyperbolic case, the situation is completely opposite to spherical space. However,
in flat space, there exists only stable black holes. Unlike to flat and hyperbolic spaces, there is not any first-order correction for internal energy in spherical space.
In spherical space, the Gibbs free energy   takes positive value only for smaller black hole.
On the other hand, in flat and hyperbolic spaces, the  Gibbs free energy takes negative values only.

Rest of the paper is organized as follows. Next, in the section 2, we recapitulate the basics  of HL black hole and their solutions. In section 3, we study the effects of thermal fluctuations on the thermodynamics and stability/instability of HL black hole  with LMP solution.  Here, the effects of thermal fluctuation are discussed for
three types of space, namely, spherical space, flat space and hyperbolic space.
 Finally, in the last section, we summarize our results with brief discussion.

\section{Ho\v rava-Lifshitz Black Hole}
Here, we first recapitulate the HL black holes and discuss the   special solutions.
We start by writing the four-dimensional gravity action of HL theory  as follow \cite{12,13},
\begin{eqnarray}\label{1}
S_{HL}&=&\int  d^{4}x\sqrt{g}N\left(\frac{2}{\kappa^2}(K_{ij} K^{ij}-\lambda K^{2})+\frac{\kappa^2 \mu^2(\Lambda_W
R-3\Lambda_W^2)}{8(1-3\lambda)}+\frac{\kappa^2\mu^2(1-4\lambda)}{32(1-3\lambda)}R^{2}
\right.\nonumber\\
&-& \left. \frac{\kappa^2\mu^2}{8}R_{ij}R^{ij}
-\frac{\kappa^2\mu}{2\omega^2}\epsilon^{ijk} R_{il}\nabla_j
R_k^{l}+ \frac{\kappa^2}{2\omega^4}C_{ij}C^{ij}\right),
\end{eqnarray}
where $\kappa^2$, $\lambda$, $\omega$, $\Lambda_W$ and $\mu$ are constant parameters,   $C^{ij}=\epsilon^{ikl} \nabla_k (R_{l}^{j} - \frac{1}{4} R\delta_{l}^{j})$ is cotton tensor,   $K_{ij}=\frac{1}{2N}(\dot{g}_{ij} -\nabla_i N_j - \nabla_j N_i )$ is extrinsic curvature written in terms of shift function $N_i$   and lapse function $N$. The cosmological constant
$\Lambda$ is related to constant parameter $\Lambda_W$  as following \cite{12}:
\begin{equation}\label{4}
\Lambda=\frac{3}{2}\Lambda_{W}.
\end{equation}
In order to study the static and spherically symmetric solution, we assume the following metric ansatz \cite{12,13}:
\begin{equation}\label{5}
ds^2=f(r)dt^2 - f^{-1}(r)dr^2 - r^{2}d\Omega^{2},
\end{equation}
where   metric of a two-dimensional   symmetric space is given by
\begin{eqnarray}\label{6}
d\Omega^{2}\equiv\left\{\begin{array}{ccc}
d\theta^2 + \sin^{2}\theta d\varphi^2 \hspace{1cm} (k=1)\\
d\theta^2 + \theta^2 d\varphi^2\hspace{1.8cm}(k=0)\\
d\theta^2 + \sinh^{2}\theta d\varphi^2\hspace{1cm}(k=-1)\\
\end{array}\right.
\end{eqnarray}
and $k$ is the curvature corresponding to spherical, flat or hyperbolic horizon, respectively.
The function $f(r)$ has the following expression for $\lambda=1$ case:
\begin{equation}\label{7}
f(r)=k + (\omega - \Lambda_W)r^{2} -\sqrt{(r(\omega(\omega - 2\Lambda_W)r^{3} + \beta))},
\end{equation}
here $\beta$ refers to an integration constant. The solution (\ref{7}) is obtained
from the equations of motion of the action (\ref{1})  with metric (\ref{5}).
There exist different solutions with different cases. For example,
in the case of $\Lambda_W=0$ and $\beta=4\omega M$, we have Kehagius-Sfetsos (KS) solution \cite{19},
\begin{equation}\label{8}
f (r)=k +\omega r^{2}- \omega r^{2}\sqrt{1+\frac{4M}{\omega r^{3}}},
\end{equation}
while in the case of $\omega=0$ and $\beta=-\frac{\gamma^2}{\Lambda_W}$, we have LMP  solution \cite{6, 12},
\begin{equation}\label{9}
f (r)=k - \Lambda_Wr^{2} - \gamma \sqrt{\frac{r}{-\Lambda_W}},
\end{equation}
which was given by Ref. \cite{6} for any $k$,  and Ref. \cite{12} for $k=1$. Here, parameter $\gamma(=aM)$ is related to the
black hole mass $M$.

The black hole thermodynamics in KS solution of HL Gravity have been studied by the Ref. \cite{18}, and thermodynamical quantities of
LMP solution of HL black hole for three different cases of spherical, flat and hyperbolic spaces are discussed in Ref. \cite{main}.
Next, we analyze the effects of thermal fluctuations
on the thermodynamics of the system.

\section{Effects of thermal fluctuations}
In this section, we discuss the effects of thermal fluctuation on
the thermodynamics of the system.
Due to the thermal fluctuations the entropy of a black hole gets   logarithmic correction, and it is interpreted as quantum effects because these are important as the size of black hole reduces due to the Hawking radiation. It is also possible to relate the black hole microscopic degrees of freedom to a conformal field theory. In this case the modular invariance of the partition function can constrain the entropy of the HL black hole.

The partition function for the HL black hole considered as the statistical mechanics
of   $N$ particles with energy spectrum $E$ is given by \cite{hawk, hawk1},
\begin{equation}\label{53}
Z = \int_0^\infty  dE \, \,  \rho (E) e^{-\beta E},
\end{equation}
where $\beta$ is  inverse of Hawking temperature $T_H$ in the units of Boltzmann constant and $\rho (E)$ is
the canonical density of the system with energy average $E$ defined as
\begin{eqnarray}\label{54}
\rho (E) = \frac{1}{2 \pi i} \int^{\beta_0+ i\infty}_{\beta_0 - i\infty}  d \beta \, \, e ^{S(\beta)}.
\end{eqnarray}
Here  entropy, $S = \beta  E   + \ln Z$, is measured around the equilibrium $\beta_0$, and all thermal fluctuations are
neglected. However, it is possible to consider mentioned thermal fluctuations and expand $S(\beta)$ around the equilibrium temperature $\beta_0$ to the first order,
\begin{equation}\label{fluc}
S = S_0 + \frac{1}{2}(\beta - \beta_0)^2 \left(\frac{\partial^2 S(\beta)}{\partial \beta^2 }\right)_{\beta = \beta_0},
\end{equation}
where $S_0$ is uncorrected entropy given by the equation (\ref{00}) and the higher order corrections of the entropy neglected.
 Following the Ref. \cite{l1}, one can write the corrected form of the entropy as
\begin{equation}\label{correctedS}
S = S_{0} + \alpha \ln (S_{0} T_H^{2}).
\end{equation}
The case of $\alpha=-\frac{1}{2}$ gives the equation (\ref{002}) hence it is generalization of the logarithmic corrected entropy.\\
The first-law of thermodynamics  is given by
\begin{equation}\label{12}
dM =T_HdS+VdP,
\end{equation}
where  $M, T_H, V$ and $P$ denote the mass, Hawking temperature, volume and pressure of
the HL black hole.
The  Hawking temperature for the black hole can be obtained from following relation:
\begin{eqnarray}\label{13}
 T_{H}&=& \frac{1}{4\pi}\left(\frac{\partial
f}{\partial r}\right)_{r=r_h},
\end{eqnarray}
where $r_{h}$ is the black hole horizon radius obtained from $f(r)=0$.
The zeroth-order entropy of the black hole is given by
\begin{equation}\label{14}
S_0=\int \frac{1}{T_H}\frac{\partial H}{\partial r_h} dr_h,
\end{equation}
where $H$ denotes the enthalpy and interpreted as the black hole mass ($H=M$) \cite{22}.
It is easy to check that $T=T_{H}$.
Thermodynamical quantities of the HL black hole for LMP solution are different for spherical space $(k=1)$, flat space $(k=0)$, and hyperbolic space $(k=-1)$.

Now, we can investigate of such correction on the thermodynamics quantities.
\subsection{Spherical space}
For the spherical space $(k=1)$,
the black hole mass is given by,
\begin{equation}\label{22}
M=a^{-1}\sqrt{\frac{-\Lambda_W}{r_h}}(1 -\Lambda_W r_h ^{2}),
\end{equation}
here  condition $f(r)|_{r=r_h}=0$ from (\ref{9})   is utilized.
Exploiting relation (\ref{13}) together with (\ref{9}) and (\ref{22}) yields the expression for
Hawking temperature,
\begin{equation}\label{24}
T_H=\frac{1}{8\pi r_h}[-1-3\Lambda_W r_h ^2].
\end{equation}
It is evident here  that  the black hole temperature depends on the   cosmological constant.
With the help of relations (\ref{14}), (\ref{22}) and (\ref{24}), the zeroth-order entropy of the black hole in spherical space is computed as
\begin{equation}\label{23}
S_{0}=\frac{8\pi}{a}\sqrt{-\Lambda_W r_h}.
\end{equation}
Now, from relation (\ref{correctedS}), it is easy to compute the first-order corrected entropy due to thermal fluctuations  as
 \begin{eqnarray}
 S=\frac{8\pi}{a}\sqrt{-\Lambda_W r_h} +\alpha \log \left[\frac{1}{8\pi a r_h}\sqrt{\frac{-\Lambda_W}{r_h}}(1+3\Lambda_Wr_h^2)^2 \right].
 \end{eqnarray}
  The expression for first-order corrected pressure  for HL black hole ($P=\frac{1}{2}T_HS $
   \cite{am,ba}) is given by
 \begin{eqnarray}
 P&=&
  -\frac{1}{2ar_h}\sqrt{-\Lambda_W r_h}(1+3\Lambda_Wr_h^2) \nonumber\\
  & -&\frac{\alpha}{16\pi r_h}(1+3\Lambda_Wr_h^2) \log \left[\frac{1}{8\pi a r_h}\sqrt{\frac{-\Lambda_W}{r_h}}(1+3\Lambda_Wr_h^2)^2 \right].
 \end{eqnarray}
 For  $\Lambda_{W}=-2$, this further reduces to
 \begin{eqnarray}\label{64}
 P&=&
  -\frac{1}{ a\sqrt{2 r_h}}(1-6r_h^2) +P_1(\alpha),
 \end{eqnarray}
 where,
 \begin{eqnarray}\label{65}
 P_1(\alpha)=
   - \frac{\alpha}{16\pi r_h}(1-6r_h^2) \log \left[\frac{1}{8\pi a r_h}\sqrt{\frac{2}{r_h}}(1-6r_h^2)^2 \right].
 \end{eqnarray}
 \begin{figure}[h!]
 \begin{center}$
 \begin{array}{cccc}
\includegraphics[width=75 mm]{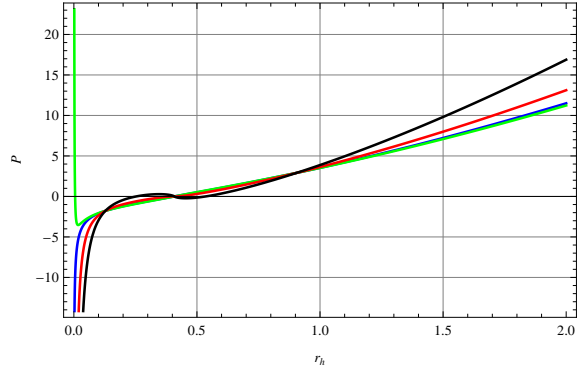}
 \end{array}$
 \end{center}
\caption{Pressure in spherical space in terms of the black hole horizon for $a=1$. Here, $\alpha=0$ denoted by blue line, $\alpha=-0.5$ denoted by green  line, $\alpha=3$ denoted by red line, and $\alpha=10$ denoted by black line.}
 \label{fig1}
\end{figure}
In Fig. \ref{fig1}, we can see the effect of thermal fluctuation  on the pressure of HL black hole in spherical space.  We notice that the pressure is negative when
horizon radius tends to smaller value for both without correction term and with positive value of
correction parameter $\alpha$. However, we see that the negative value of $\alpha$ compensates
the negative pressure and makes it positive even for the smaller  horizon radius.
We found that the pressure takes asymptotic values for point-like back hole. After a
critical value, the pressure increases with horizon radius and  increases more sharply
with higher values of $\alpha$.

 The corrected volume of black hole thus is given by,
\begin{eqnarray}\label{26}
V&=& \left(\frac{\partial H}{\partial P}\right)_S=\left(\frac{\partial M}{\partial r_h}\frac{\partial r_h}{\partial P}\right)_S,\nonumber\\
&=&\frac{4 \pi   r^2\left(3 \Lambda  r^2+1\right)}{\Lambda  \left(3 \Lambda  r^2-1\right)}
\left(-\frac{\Lambda }{r}\right)^{3/2}\left(8 \pi  r \sqrt{-\frac{\Lambda }{r}} +a \alpha  \log \left(\frac{\sqrt{-\frac{\Lambda }{r}} \left(3 \Lambda  r^2+1\right)^2}{8 \pi  a r}\right)
 \right)^{-1}.
\end{eqnarray}
This further simplifies to
\begin{eqnarray}\label{261}
V &=& \frac{1}{2}\left(\frac{1+3\Lambda_W r_h^2}{1-3\Lambda_W r_h^2} \right) -\frac{\alpha a}{16\pi\sqrt{-\Lambda_W r_h}}\left(\frac{1+3\Lambda_W r_h^2}{1-3\Lambda_W r_h^2} \right)
\log \left(\frac{\sqrt{- {\Lambda } } \left(3 \Lambda  r^2+1\right)^2}{8 \pi  a r^{3/2}}\right).
\end{eqnarray}
  The first-order corrected HL black hole Helmholtz free energy  is derived as
 \begin{eqnarray}
 F&=&-\int S dT_H,\nonumber\\
 &=&\frac{2}{a}\sqrt{\frac{-\Lambda_W}{r_h}}(1+\Lambda_W r_h^2) -\frac{3\alpha}{16\pi r_h}(1+5\Lambda_W r_h^2) \nonumber\\
 &+&\alpha\frac{(1+3\Lambda_W r_h^2)}{8\pi r_h}\log\left[\frac{1}{8\pi a r_h}\sqrt{\frac{-\Lambda_W}{r_h}}(1+3\Lambda_Wr_h^2)^2 \right].
 \end{eqnarray}
 Now,  by setting $\Lambda_{W}=-2$, then  the  Helmholtz free energy reduces to
 \begin{eqnarray}
 F
 &=&\frac{2}{a}\sqrt{\frac{2}{r_h}}(1-2 r_h^2)+F_1(\alpha)+F_2(\alpha),
 \end{eqnarray}
 where
 \begin{eqnarray}
 F_1(\alpha)  &=&-\frac{3\alpha}{16\pi r_h}(1-10 r_h^2), \nonumber\\
 F_2(\alpha)  &=&\alpha\frac{(1-6 r_h^2)}{8\pi r_h}\log\left[\frac{1}{8\pi a r_h}\sqrt{\frac{2}{r_h}}(1-6 r_h^2)^2 \right].
 \end{eqnarray}
 \begin{figure}[h!]
 \begin{center}$
 \begin{array}{cccc}
\includegraphics[width=75 mm]{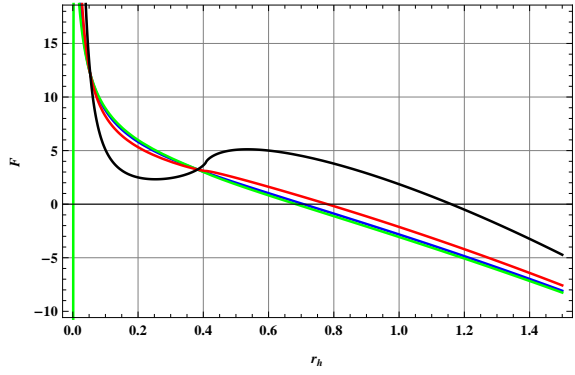}
 \end{array}$
 \end{center}
\caption{Helmholtz free energy in spherical space in terms of the black hole horizon for $a=1$. Here, $\alpha=0$ denoted by blue line, $\alpha=-0.5$ denoted by green  line, $\alpha=3$ denoted by red line, and $\alpha=10$ denoted by black line.}
 \label{fig2}
\end{figure}
In the plot \ref{fig2}, the change in behavior of the Helmholtz free energy due to the logarithmic correction is shown. We examined both positive and negative value of correction coefficient. In the case of negative $\alpha$, the Helmholtz free energy takes negative
 asymptotic value when horizon radius tends to zero. However, uncorrected and corrected Helmholtz free energy with positive $\alpha$ do not fall asymptotically near the vanishing
 small $r_h$. In comparison to smaller $\alpha$, the higher  $\alpha$ effects Helmholtz free energy   differently and the positive span of Helmholtz free energy for higher $\alpha$ is more.
   For the very large $r_{h}$,  there is no main differences in corrected and uncorrected energy. There exists a critical radius of horizon for which  the Helmholtz free energy is constant irrespective to all different cases.

Now, in order to study the critical points and  stability, we calculate specific heat as
\begin{eqnarray}\label{29}
C &=& T_H\frac{\partial S}{\partial T_H},\nonumber\\
&=&-\frac{4\pi}{a}\sqrt{-\Lambda_W r_h}\left(\frac{1+3\Lambda_W r_h^2}{1-3\Lambda_W r_h^2}\right)
+\frac{3\alpha}{2}\left(\frac{ 1-5\Lambda_W r_h^2}{1+3\Lambda_W r_h^2}\right).
\end{eqnarray}
\begin{figure}[h!]
 \begin{center}$
 \begin{array}{cccc}
\includegraphics[width=75 mm]{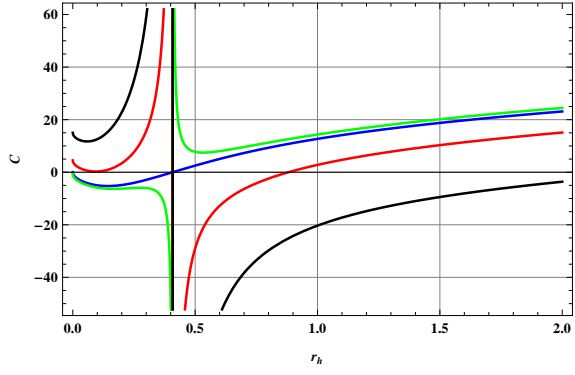}
 \end{array}$
 \end{center}
\caption{Specific heat in spherical space in terms of the black hole horizon for $\Lambda_W=-2$ and $a=1$.
Here, $\alpha=0$ denoted by blue line, $\alpha=-0.5$ denoted by green  line, $\alpha=3$ denoted by red line, and $\alpha=10$ denoted by black   line.}
 \label{fig3}
\end{figure}
 Fig. \ref{fig3} discusses some important aspects about the stability/instability of
 HL black hole in spherical space. We find that there exists a critical radius $r_{hc}$ for HL black hole and the large black hole ($r_h > r_{hc}$)
 without thermal fluctuations is found in completely stable phase, however
 the small black hole ($r_h < r_{hc}$)  exists in completely unstable phase. Remarkably, we notice that there exists phase transition at critical radius due to the thermal fluctuations.
 Also, with the positive coefficient $\alpha$, we can  obtain some stable regions for the small HL black hole in spherical space. The correction with the positive  $\alpha$ also causes
  some instabilities for  larger black hole as well.

Finally the internal energy is obtained as,
\begin{equation}\label{30}
E=M-PV=a^{-1}\sqrt{\frac{2}{r_h}}(1 +2 r_h ^{2})+a^{-1}\frac{1}{4}\sqrt{\frac{2}{r_h}}\frac{(1-6r^2_h)^2}{(1+6r^2_h)}.
\end{equation}
From the expression, it is evident that there is no first-order correction on
internal energy in spherical space. However, still there is possibility of  higher-order
corrections on the internal energy.
The Gibbs free energy using the relation
\begin{eqnarray}
G&=&M-T_HS,\nonumber\\
&=&\frac{2}{a} \sqrt{\frac{2}{r_h}}(1 -2 r_h ^{2})+\frac{\alpha}{8\pi r_h}
(1-6 r_h^2)\log\left[\frac{1}{4\sqrt{2}\pi a r_h^{3/2}} (1-6r_h^2)^2 \right],
\end{eqnarray}
where $\Lambda_{W}=-2$ is set.
\begin{figure}[h!]
 \begin{center}$
 \begin{array}{cccc}
\includegraphics[width=75 mm]{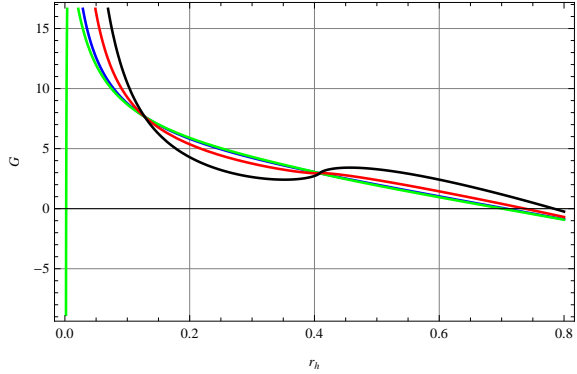}
 \end{array}$
 \end{center}
\caption{Gibbs free energy in spherical space in terms of the black hole horizon for   $a=1$. Here,
 $\alpha=0$ denoted by blue line, $\alpha=-0.5$ denoted by green  line, $\alpha=3$ denoted by red line, and $\alpha=10$ denoted by black   line.}
 \label{fig4}
\end{figure}
In Fig. \ref{fig4}, we see that both uncorrected and corrected  Gibbs free energy  take  positive values for smaller black hole. However, corrected Gibbs free energy  with  negative $\alpha$ takes negative asymptotic value for point like  back holes.
  There exists two critical points and  Gibbs free energy  is positive in between there two points. However, for large sized black hole, there is not much effect of thermal fluctuation
on  Gibbs free energy  and it becomes more negative with larger radius.

\subsection{Flat space}
In order to discuss the
 first-order corrected thermodynamics quantities for flat space
$(k=0)$, the  LMP  solution is given by
\begin{equation} \label{ff}
f (r)=  - \Lambda_Wr^{2} - aM \sqrt{\frac{r}{-\Lambda_W}},
\end{equation}
which leads to following expression for mass:
\begin{equation}\label{32}
M=\frac{1}{a}(-\Lambda_W r_h)^\frac{3}{2}.
\end{equation}
Here we see that this is an increasing function of $r_{h}$.
Now, utilizing relations (\ref{13}), (\ref{ff}) and (\ref{32}), we are able to calculate the Hawking temperature
\begin{equation}\label{35}
T_H=-\frac{3}{8\pi} \Lambda_W r_h.
\end{equation}
Here, we see that the magnitude of temperature of black hole increases linearly with
the cosmological constant.

Now, the zeroth-order entropy for the black hole is calculated as
\begin{equation}\label{34}
S_{0}=\frac{8\pi}{a}\sqrt{-\Lambda_W r_h},
\end{equation}
where (\ref{14}), (\ref{32}) and (\ref{35}) are utilized.
 With the help of (\ref{correctedS}), it matter of calculation only to write
  the first-order corrected entropy due to thermal fluctuations  as
\begin{equation}\label{ss}
S =\frac{8\pi}{a}\sqrt{-\Lambda_W r_h} +\alpha \log \left[\frac{9}{8\pi a}(-\Lambda_W r_h)^{5/2}\right].
\end{equation}
Following the procedure of above section, the first-order corrected  pressure is given by
\begin{equation}\label{36}
P= \frac{3}{2a}(-\Lambda_W r_h)^{3/2}-\frac{3\alpha}{16\pi} \Lambda_W r_h\log \left[\frac{9}{8\pi a}(-\Lambda_W r_h)^{5/2}\right].
\end{equation}
 \begin{figure}[h!]
 \begin{center}$
 \begin{array}{cccc}
\includegraphics[width=75 mm]{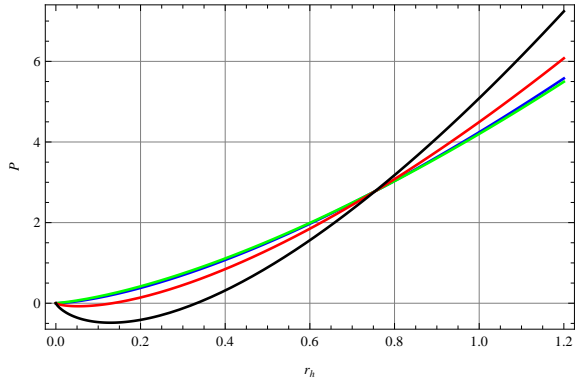}
 \end{array}$
 \end{center}
\caption{Pressure in flat space in terms of the black hole horizon for $a=1$ and $\Lambda_W=-2$. Here, $\alpha=0$ denoted by blue line, $\alpha=-0.5$ denoted by green  line, $\alpha=3$ denoted by red line, and $\alpha=10$ denoted by black line.}
 \label{fig5}
\end{figure}
From the plot of Fig. \ref{fig5}, we see that for smaller black holes in flat space the pressure becomes more negative when the value of correction parameter $\alpha$ takes large positive values. However, without correction or with correction with negative $\alpha$, the pressure
is positive always. But after the critical value, pressure   increases more sharply with
size of black holes in case of  positive $\alpha$.

With the help of expression for pressure, one can obtain the corrected volume
in case of flat space as,
\begin{equation}\label{37}
V= \frac{2}{3}-\frac{\alpha a}{36\pi\sqrt{-\Lambda_W r_h}}\left( 5+\log \left[\frac{9}{8\pi a}(-\Lambda_W r_h)^{5/2}\right]\right).
\end{equation}

In this case, the first-order corrected black hole Helmholtz free energy is obtained using the relations (\ref{35})  and (\ref{ss}), as following:
\begin{eqnarray}\label{66}
F= \frac{2}{a}(-\Lambda_W r_h)^{3/2}-\frac{15\alpha}{16\pi}\Lambda_W r_h +\frac{3\alpha}{8\pi}\Lambda_W r_h \log \left[\frac{9}{8\pi a}(-\Lambda_W r_h)^{5/2}\right].
\end{eqnarray}
 It is clear that the uncorrected Helmholtz free energy is twice to mass $M$ (\ref{32}).
 \begin{figure}[h!]
 \begin{center}$
 \begin{array}{cccc}
\includegraphics[width=75 mm]{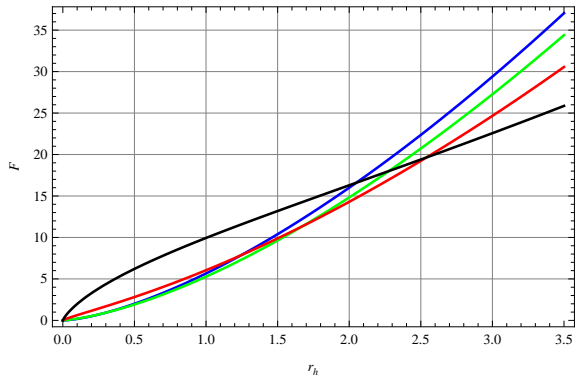}
 \end{array}$
 \end{center}
\caption{Helmholtz free energy in flat space in terms of the black hole horizon for $a=1$ and $\Lambda_W=-2$. Here, $\alpha=0$ denoted by blue line, $\alpha=-0.5$ denoted by green  line, $\alpha=3$ denoted by red line, and $\alpha=10$ denoted by black line.}
 \label{fig6}
\end{figure}
 From Fig. \ref{fig6}, it is evident that the Helmholtz free energy for black hole in flat space is  an increasing function with $r_h$.  We observe that for smaller value of
 $\alpha$ the  corrected Helmholtz free energy behaves more or less like uncorrected one.
 However, for rather higher value of $\alpha$, the  Helmholtz free energy starts behaving
  differently and increases  with lesser slope.

The heat capacity is an important parameter to study stability of the black hole.
So, for flat space, we found the following expression:
\begin{equation}\label{40}
C = \frac{4\pi}{a}\sqrt{-\Lambda_W r_h}+\frac{5}{2}\alpha,
\end{equation}
 \begin{figure}[h!]
 \begin{center}$
 \begin{array}{cccc}
\includegraphics[width=75 mm]{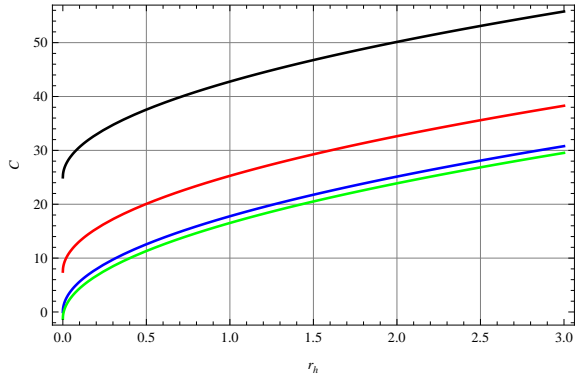}
 \end{array}$
 \end{center}
\caption{Heat capacity in flat space in terms of the black hole horizon for $a=1$ and $\Lambda_W=-2$. Here, $\alpha=0$ denoted by blue line, $\alpha=-0.5$ denoted by green  line, $\alpha=3$ denoted by red line, and $\alpha=10$ denoted by black line.}
 \label{fig7}
\end{figure}
We can see that uncorrected heat capacity $C_{0}=S_{0}/2$ and this justifies the corrected relation (\ref{002}).
From the both expression (\ref{40}) and Fig. \ref{7}, we see that the heat capacity  in flat space is always
positive with the negative cosmological constant. Therefore, there exists
stable black holes only.

Now, the internal energy ($E=M-PV$) is obtained
as follow,
\begin{equation}\label{41}
E=-\frac{5\alpha}{24\pi}\Lambda_W r_h +\frac{\alpha}{12\pi}\Lambda_W r_h\log \left[\frac{9}{8\pi a}(-\Lambda_W r_h)^{5/2}\right].
\end{equation}
Here we see that the internal energy for black hole depends  completely   on logarithmic correction. There is no zeroth-order  internal energy for black hole in flat space.
\begin{figure}[h!]
 \begin{center}$
 \begin{array}{cccc}
\includegraphics[width=75 mm]{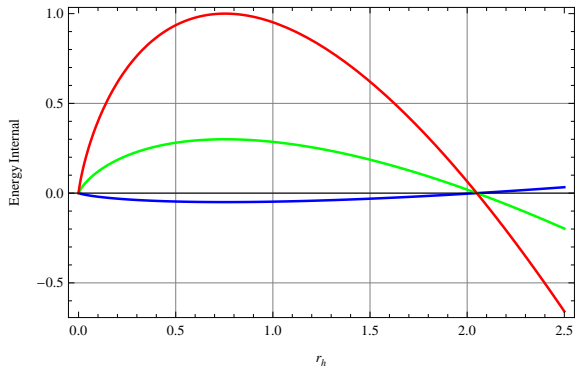}
 \end{array}$
 \end{center}
\caption{Internal energy in flat space in terms of the black hole horizon for $a=1$ and $\Lambda_W=-2$. Here, $\alpha=-0.5$ denoted by blue line, $\alpha=3$ denoted by green  line, $\alpha=10$ denoted by red line.}
 \label{fig8}
\end{figure}
In Fig. \ref{fig8}, for negative $\alpha$, the internal energy is negative and positive
for small black hole (smaller than critical radius) and large black hole (larger than critical radius), respectively. However, for positive $\alpha$, the internal energy is positive and negative for small black hole  and large black hole, respectively.

  Finally,  we derive the corrected  Gibbs free energy for LMP black hole in flat space as,
\begin{equation}\label{70}
G= -\frac{2}{a}(-\Lambda_W r_h)^{3/2}+\frac{3\alpha}{8\pi}\Lambda_W r_h \log \left[\frac{9}{8\pi a}(-\Lambda_W r_h)^{5/2}\right].
\end{equation}
Here we observe that the zeroth-order Gibbs free energy is negative to zeroth-order  black   Helmholtz free energy.
\begin{figure}[h!]
 \begin{center}$
 \begin{array}{cccc}
\includegraphics[width=75 mm]{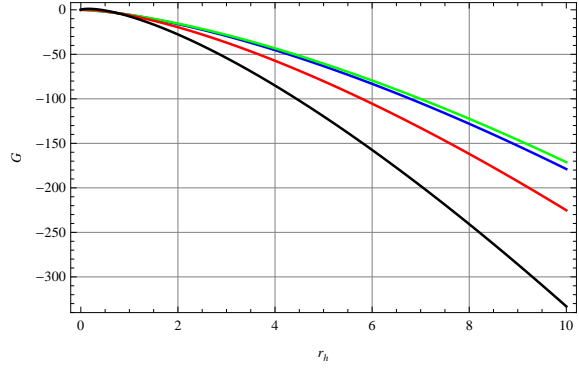}
 \end{array}$
 \end{center}
\caption{Gibbs free energy in flat space in terms of the black hole horizon for $a=1$ and $\Lambda_W=-2$. Here, $\alpha=0$ denoted by blue line, $\alpha=-0.5$ denoted by green  line, $\alpha=3$ denoted by red line, and $\alpha=10$ denoted by black line.}
 \label{fig9}
\end{figure}
In Fig. \ref{fig9}, we see that the Gibbs free energy in flat space is negative
in all cases. Although, the higher positive values make  the Gibbs free energy more negative,
the behavior of Gibbs free energy does not change on the correction parameter.

\subsection{Hyperbolic space}
In this subsection, we study the effects of  quantum gravity effects on
LMP black hole in hyperbolic
space $(k=-1)$. In this case,
the  LMP  solution is given by
\begin{equation} \label{ff1}
f (r)= -1 - \Lambda_Wr^{2} - aM \sqrt{\frac{r}{-\Lambda_W}}.
\end{equation}
The black hole mass at horizon $f (r)|_{r=r_h}=0$ is calculated by,
\begin{equation}\label{43}
M=H=\frac{1}{a}\sqrt{\frac{-\Lambda_W}{r_h}} [-1-\Lambda_W
r_h^2].
\end{equation}
It can be seen that the magnitude of the cosmological constant increases the
black hole mass.

Exploiting relation (\ref{13}), the Hawking  temperature in this case has following expression:
\begin{equation}\label{45}
T_H=\frac{1}{8\pi r_h}(1-3\Lambda_W r_h^2).
\end{equation}
The expression (\ref{14}), (\ref{43}) and (\ref{45}) induce the following zeroth-order entropy of the black hole in hyperbolic  space:
\begin{equation}\label{44}
S_{0}=\frac{8\pi}{a}\sqrt{-\Lambda_W r_h},
\end{equation}
here we notice that the zeroth-order entropy for both the spherical and hyperbolic
spaces is same.
Utilizing  relation (\ref{correctedS}),  the first-order corrected entropy  in hyperbolic space is given by
 \begin{eqnarray}
 S=\frac{8\pi}{a}\sqrt{-\Lambda_W r_h} +\alpha \log \left[\frac{1}{8\pi a r_h}\sqrt{\frac{-\Lambda_W}{r_h}}(1-3\Lambda_Wr_h^2)^2 \right].
 \end{eqnarray}
It is clear that correction terms are different for the cases of the spherical and hyperbolic
spaces.
 Now, one can find pressure
as follow,
\begin{eqnarray}\label{46}
 P&=& \frac{1}{2}T_H S=
   \frac{1}{2ar_h}\sqrt{-\Lambda_W r_h}(1-3\Lambda_Wr_h^2) \nonumber\\
  & +&\frac{\alpha}{16\pi r_h}(1-3\Lambda_Wr_h^2) \log \left[\frac{1}{8\pi a r_h}\sqrt{\frac{-\Lambda_W}{r_h}}(1-3\Lambda_Wr_h^2)^2 \right],
 \end{eqnarray}
which is positive for all values of negative cosmological constant.
\begin{figure}[h!]
 \begin{center}$
 \begin{array}{cccc}
\includegraphics[width=75 mm]{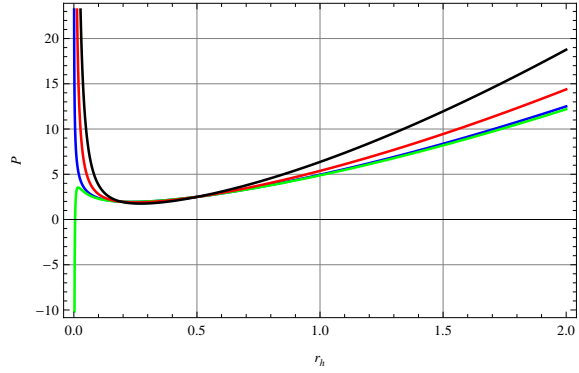}
 \end{array}$
 \end{center}
\caption{Pressure in hyperbolic space in terms of the black hole horizon for $a=1$ and $\Lambda_W=-2$. Here, $\alpha=0$ denoted by blue line, $\alpha=-0.5$ denoted by green  line, $\alpha=3$ denoted by red line, and $\alpha=10$ denoted by black line.}
 \label{fig10}
\end{figure}
In Fig. \ref{fig10}, we can see the effect of thermal fluctuation  on the pressure of HL black hole in hyperbolic space.  We find that for both the cases of
corrected and uncorrected one, the pressure is increasing function and takes positive value for finite size of black hole. However, the pressure is decreasing function for the
uncorrected and corrected with positive $\alpha$ for  vanishingly small $r_h$.
In case of negative $\alpha$, the pressure becomes asymptotically
negative for $r_h$ tending to zero.

Once expression for pressure is know, it is easy to compute volume  with the help of relation $V= \left(\frac{\partial M}{\partial P}\right)_S$ as
\begin{eqnarray}\label{47}
V&=& 2\left(\frac{1-3\Lambda_W r_h^2}{1-9\Lambda_Wr_h^2} \right)+\frac{2\alpha a}{3\pi\sqrt{-\Lambda_W r_h}}\frac{(1+3\Lambda_W r_h^2)(1+5\Lambda_W r_h^2)}{(1-9\Lambda_W r_h^2)^2} \nonumber\\
&+& \frac{4\alpha a}{9\pi\sqrt{-\Lambda_W r_h}} \left(\frac{ 1+3\Lambda_W r_h^2  }{ 1-9\Lambda_W r_h^2 }\right)^2 \log  \left[\frac{1}{8\pi a r_h}\sqrt{\frac{-\Lambda_W}{r_h}}(1-3\Lambda_Wr_h^2)^2 \right].
\end{eqnarray}
In the case of hyperbolic space $(k=-1)$, the black hole Helmholtz free energy obtained using the relation, $F=-\int SdT_H$, as
\begin{equation}\label{71}
F=- \frac{2}{a}\sqrt{\frac{-\Lambda_W}{r}}(1-\Lambda_W r_h^2)+F_{1}(\alpha)+F_{2}(\alpha),
\end{equation}
where
\begin{eqnarray}\label{72}
F_{1}(\alpha)&=&\frac{3\alpha}{16\pi r_h}(1-5\Lambda r_h^2),\nonumber\\
F_{2}(\alpha)&=& -\frac{\alpha}{8\pi r_h}(1-3\Lambda_W r_h^2)\log  \left[\frac{1}{8\pi a r_h}\sqrt{\frac{-\Lambda_W}{r_h}}(1-3\Lambda_Wr_h^2)^2 \right].
\end{eqnarray}
If  we set $\Lambda_{W}=-2$, this reduces to
\begin{eqnarray}\label{72}
F &=&- \frac{2\sqrt{2}}{a\sqrt{r_h}} (1+2 r_h^2)+\frac{3\alpha(1+10 r_h^2)}{16\pi r_h}  -\frac{\alpha(1+6 r_h^2)}{8\pi r_h}\log  \left[\frac{1}{2\sqrt{2}\pi a r_h^{3/2}} (1+6r_h^2)^2 \right].
\end{eqnarray}
\begin{figure}[h!]
 \begin{center}$
 \begin{array}{cccc}
\includegraphics[width=75 mm]{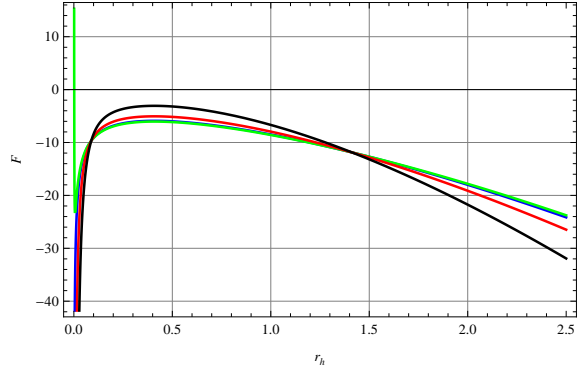}
 \end{array}$
 \end{center}
\caption{Helmholtz free energy in hyperbolic space in terms of the black hole horizon for $a=1$. Here, $\alpha=0$ denoted by blue line, $\alpha=-0.5$ denoted by green  line, $\alpha=3$ denoted by red line, and $\alpha=10$ denoted by black line.}
 \label{fig11}
\end{figure}
In Fig. \ref{fig11}, we notice that the Helmholtz free energy for finite sized  HL black hole in
hyperbolic space  is negative. However, there exist a critical point after which the
behavior of Helmholtz free energy changes. For point-like HL black hole, the
corrected Helmholtz free energy for negative $\alpha$  takes asymptotically positive value
only.  However, for the same size of black hole the corrected Helmholtz free energy
take negative value very fast analogous to uncorrected case.

The logarithmic corrected specific heat is given by
\begin{equation}\label{50}
C = -\frac{4\pi}{a}\sqrt{-\Lambda_W r_h}\left(\frac{1-3\Lambda_W r_h^2}{1+3\Lambda_W r_h^2}\right)+\frac{3\alpha}{2}\left(\frac{1+5\Lambda_W r_h^2}{1+3\Lambda_W r_h^2}\right).
\end{equation}
\begin{figure}[h!]
 \begin{center}$
 \begin{array}{cccc}
\includegraphics[width=75 mm]{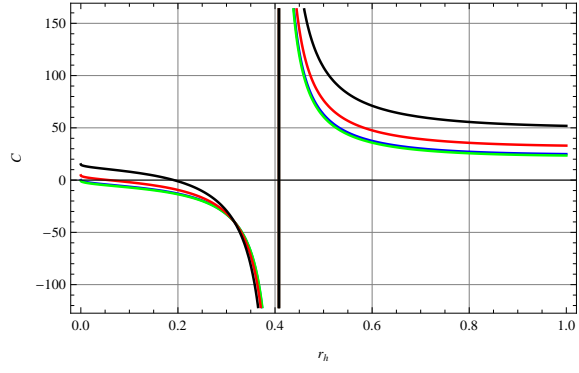}
 \end{array}$
 \end{center}
\caption{Specific heat in hyperbolic space in terms of the black hole horizon for $a=1$. Here, $\alpha=0$ denoted by blue line, $\alpha=-0.5$ denoted by green  line, $\alpha=3$ denoted by red line, and $\alpha=10$ denoted by black line.}
 \label{fig12}
\end{figure}
From the plot of Fig. \ref{fig12}, we observe that  in hyperbolic space  the phase transition
occurs
for both corrected and uncorrected cases of specific heat.
The phase transition occurs at critical radius ($r_{hc}$). Interestingly, we find that  the large HL black hole ($r_{h}>r_{hc}$) is in stable phase,  while small black hole ($r_{h}<r_{hc}$) is completely at unstable phase.
 In absence of quantum effects, there are some instabilities for the small radius. Moreover,
  in presence of thermal fluctuations with positive parameter, the black hole can have stable regions for the small HL black hole also.

Now, we calculate the first-order corrected  internal energy as follow,
\begin{eqnarray} \label{51}
E&=&-\frac{2}{a}\sqrt{\frac{-\Lambda_W}{r_h}}\left(\frac{1-7\Lambda_W r_h^2}{1-9\Lambda_W r_h^2}\right)+\frac{\alpha}{3\pi   r_h}\frac{(1-9\Lambda^2_W r_h^4)(1+5\Lambda_W r_h^2)}{(1-9\Lambda_W r_h^2)^2}\nonumber\\
&+& \frac{\alpha(1-3\Lambda_W r_h^2)(25-12\Lambda_W r_h^2+387\Lambda^2_W r_h^4)}{72\pi r_h(1-9\Lambda_W r_h^2)^2} \log  \left[\frac{\sqrt{ -\Lambda_W}}{8\pi a r^{3/2}_h}
(1-3\Lambda_Wr_h^2)^2 \right].
\end{eqnarray}
\begin{figure}[h!]
 \begin{center}$
 \begin{array}{cccc}
\includegraphics[width=75 mm]{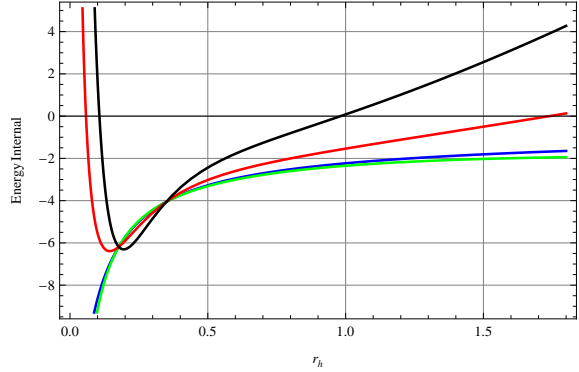}
 \end{array}$
 \end{center}
\caption{Internal energy in hyperbolic space in terms of the black hole horizon for $a=1$. Here, $\alpha=0$ denoted by blue line, $\alpha=-0.5$ denoted by green  line, $\alpha=3$ denoted by red line, and $\alpha=10$ denoted by black line.}
 \label{fig13}
\end{figure}
Due to thermal fluctuation, the internal energy shows different behavior
for small black hole ($r_h <0.4$) with positive correction coefficient.
There exist one minimum for  the case of  positive correction coefficient.
However, the corrected internal energy with negative correction coefficient
follows similar behavior to that of without correction.

Finally, the corrected Gibbs free energy for hyperbolic space is calculated by
\begin{eqnarray}
G &=&-\frac{2}{a} \sqrt{\frac{-\Lambda_W}{r_h}}(1 -\Lambda_W r_h ^{2})-\frac{\alpha}{8\pi r_h}
(1-3\Lambda_W r_h^2)\log  \left[\frac{\sqrt{-\Lambda_W}}{8\pi a r^{3/2}_h}(1-3\Lambda_Wr_h^2)^2 \right].
\end{eqnarray}
For fixed value of  $\Lambda_{W}=-2$, this turns to following:
\begin{eqnarray}
G &=&-\frac{2}{a} \sqrt{\frac{2}{r_h}}(1 +2 r_h ^{2})-\frac{\alpha}{8\pi r_h}
(1+6 r_h^2)\log  \left[\frac{1}{4\sqrt{2}\pi a r^{3/2}_h}(1+6r_h^2)^2 \right].
\end{eqnarray}
\begin{figure}[h!]
 \begin{center}$
 \begin{array}{cccc}
\includegraphics[width=75 mm]{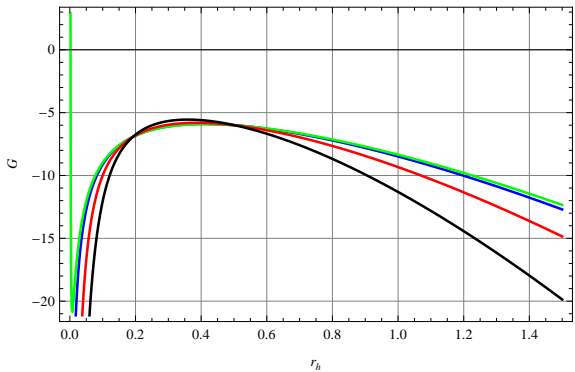}
 \end{array}$
 \end{center}
\caption{Gibbs free energy in hyperbolic space in terms of the black hole horizon for $a=1$. Here, $\alpha=0$ denoted by blue line, $\alpha=-0.5$ denoted by green  line, $\alpha=3$ denoted by red line, and $\alpha=10$ denoted by black line.}
 \label{fig14}
\end{figure}
In Fig. \ref{fig14}, we see that  there exists two critical radius for the Gibbs free energy of black hole with finite size in hyperbolic space.  The corrected Gibbs free energy  takes
positive value (asymptotically) only with negative correction coefficient. However, for
small black hole,
the corrected Gibbs free energy with positive correction coefficients show  similar
behavior to that of uncorrected case.
\section{Concluding remarks}
In this paper, we considered the LMP solution of HL black hole in flat, spherical and hyperbolic spaces and studied the effects of quantum correction  on the black hole thermodynamics. Quantum corrections appear due to the statistical thermal fluctuations and have logarithmic shape. Here, we  introduced the origin of logarithmic correction and finally discussed the effects of such correction on the thermodynamics quantities.

The first order-corrected equations of states are calculated so that they satisfy the
first-law of thermodynamics for all kind of curvatures.
The quantum effects on thermodynamical quantities is shown in the various
plots above. For example,
 in case of spherical space,     the pressure is found negative when horizon radius is small for both without correction   and with positive   correction
coefficients. The negative value of   correction
coefficient  compensates the negative
pressure and makes it positive even for the smaller  horizon radius.   However, in case of flat space, the pressure with large positive coefficient is  found
 negative for small black hole only.
 In case of hyperbolic space,  the pressure is an increasing
function and takes positive value for finite size of black hole.  The corrected pressure is asymptotically negative when horizon radius tends to zero. We have observed that
the Helmholtz free energy for spherical space takes negative asymptotic value for
 negative correction parameter only when horizon radius tends to zero.
  On the other hand, in flat space, the Helmholtz free energy   is an increasing function always. The behavior of the Helmholtz free energy
is found opposite to the pressure  in  case of hyperbolic space.

In spherical space, we have found the critical radius $r_{hc}$ for HL black hole and have
observed that the large black hole (with horizon radius $r_h$ greater than critical radius)
 without thermal fluctuations is in completely stable phase, however the
 small black hole ($r_h < r_{hc}$)  is in completely unstable phase. Remarkably,  the thermal fluctuations causes a phase transition at critical radius.
 Even   some stable phase exists  for the small HL black hole  with the positive coefficient in spherical space. However, the correction with the positive correction parameter  is also responsible for   some instabilities for  larger black hole.    In flat space, the black hole is always stable and thermal fluctuations do not affect the stability of black hole.
Moreover, in hyperbolic space,  the phase transition
occurs for both with and without thermal fluctuations.
The phase transition occurs at critical radius. Incidentally, we have found that  the large HL black hole  is in stable phase,  while instabilities occur for small black holes.
  The thermal fluctuations with positive parameter removes instabilities of the small HL black hole also.

Apart from flat and hyperbolic cases, we have not found any correction for internal energy  to
the first-order in spherical space.  However, the higher-order corrections may still be
present. Further, we have studied quantum effects on the Gibbs free energy. We have found that
in spherical space, the Gibbs free energy  takes only positive value   for the smaller black hole.
On the other hand, in flat and hyperbolic spaces, the  Gibbs free energy for  black hole
with finite horizon radius takes negative values only.
In hyperbolic space, the first-order corrected Gibbs free energy  with negative correction coefficient takes asymptotic  positive
value when the horizon radius tends to zero limit. However, for small black
holes, the corrected Gibbs free energy with positive correction coefficients shows similar
behavior to that of without thermal fluctuations.

\end{document}